# Scaling of Memories and Crossover in Glassy Magnets

A. M. Samarakoon[1,2,+], M. Takahashi[3,+], D. Zhang[1], J. Yang[1], N. Katayama[1,4], R. Sinclair[5], H. D. Zhou[5], S. O. Diallo[2], G. Ehlers[2], D. A. Tennant[2], S. Wakimoto[6], K. Yamada[7], G-W. Chern[1], T. J. Sato[3*], S.-H. Lee[1]*

[1] Department of Physics, University of Virginia, Charlottesville, Virginia 22904, USA

[2] Oak Ridge National Laboratory, Oak Ridge, Tennessee 37831, USA

[3] Institute of Multidisciplinary Research for Advanced Materials, Tohoku University, Katahira, Sendai 980-857, Japan

[4] Department of Applied Physics, Nagoya University, Nagoya 464-8603, Japan

[5] Department of Physics and Astronomy, University of Tennessee, Knoxville, Tennessee 37996, USA

[6] Materials Sciences Research Center, Japan Atomic Energy Agency, 2-4 Shirakata Shirane, Tokai, Naka, Ibaraki 319-1195, Japan

[7] Institute of Materials Structure Science, High Energy Accelerator Research Organization, Oho, Tsukuba 305-0801, Japan

[+] AMS and MT made equal contributions to this work.

* To whom correspondence should be addressed. E-mail: shlee@virginia.edu or taku@tagen.tohoku.ac.jp

**Abstract:** Glassiness is ubiquitous and diverse in characteristics in nature. Understanding their differences and classification remains a major scientific challenge. Here, we show that scaling of magnetic memories with time can be used to classify magnetic glassy materials into two distinct classes. The systems studied are **high temperature superconductor-related materials, spin-orbit Mott insulators, frustrated magnets, and dilute magnetic alloys. Our bulk magnetization measurements reveal that most densely populated magnets exhibit similar memory behavior characterized by a relaxation exponent of $1 - n \approx 0.6(1)$. This exponent is different from $1 - n \approx 1/3$ of dilute magnetic alloys that was ascribed to their hierarchical and fractal energy landscape, and is also different from $1 - n = 1$ of the conventional Debye relaxation expected for a spin solid, a state with long range order. Furthermore, our systematic study on dilute magnetic alloys with varying magnetic concentration exhibits crossovers among the two glassy states and spin solid.**

Magnetic glassy systems present a unique opportunity for searching possible universal phenomena associated with glassy behaviors. This is because glass phase exists in a wide range of magnetic materials that are described by seemingly very different spin interactions. The most well-known common features of the magnetic glassy behaviors are the lack of long range magnetic order and the field-cooled (FC) and zero-field-cooled (ZFC) hysteresis found in the bulk susceptibility[1,2]. The term spin glass was coined in 1970s to describe the low temperature behaviors of dilute magnetic alloys that are made of nonmagnetic metals with low concentrations of magnetic impurities[1,3]. The canonical glassy behaviors are manifested in intriguing phenomena called aging, rejuvenation, and memory effects[4]. While aging simply refers to the time-span dependence of relaxation phenomena in the glassy state, rejuvenation describes the re-thermalization whenever the system is further cooled after waiting at some temperature. The states accessed while aging can be retrieved upon re-heating, which is called memory effect. Several theories have been proposed to understand the physics of the spin glass.

Various systems other than the dilute magnetic alloys also exhibit the aforementioned characteristic glassy behaviors at low temperatures, even when the magnetic moments are densely populated. For example, glassy behaviors have been observed in the phase diagrams of high temperature superconducting materials, cuprates[5,6] and iron-based superconductors[7]. Another example is the so-called spin-orbit Mott insulators, $Li_2RhO_3$[8,9] and $Na_2Ir_{1-x}Ti_xO_3$[10], which exhibit anisotropic Kitaev-type exchange interactions. Yet another is a set of geometrically frustrated magnets, pyrochlores such as $Y_2Mo_2O_7$[11], spinels such as $ZnFe_2O_4$[12], and the quasi-two-dimensional bi-pyramid compounds $SrCr_{9p}Ga_{12-9p}O_{19}$ (SCGO)[13-17] and $BaCr_{9p}Ga_{12-9p}O_{19}$ (BCGO)[18]. We emphasize that the magnetic interactions of these systems seem to be quite different in nature. For instance, the parent compound of high-$T_c$ superconductors $La_{2-x}Sr_xCuO_4$ (LSCO) is a Mott insulator with a conventional Neel spin order[19]. The entire magnetic excitation spectrum of $La_2CuO_4$ can be understood by an effective spin Hamiltonian with dominant nearest neighbor antiferromagnetic coupling constant $J = 104$ meV[20]. The iron chalcogenide $Fe_{1+y}Te$ displays a bi-collinear antiferromagnetic stripe order[21,22]. Magnetic interactions in the two spin-orbit Mott insulators, $Li_2RhO_3$ and $Na_2Ir_{1-x}Ti_xO_3$ are dominated by highly anisotropic Kitaev exchange couplings[23,24]. Remarkably, despite their different nature of magnetic interactions, all the systems show the same FC-ZFC hysteresis at low temperatures. A natural question to ask is whether or not there is a unifying concept that can unite and also classify these various glassy magnets.

Here, we address this issue by investigating memory effects of several of the aforementioned exemplary systems using the bulk magnetization measurements. We performed thermo-remanent magnetization (TRM)[25-27] measurements on five different compounds, which can be divided into three categories: (1) the high temperature superconducting materials, cuprates and Fe-chalcogenides, (2) Kitaev-model-related systems $Li_2RhO_3$ and $Na_2Ir_{1-x}Ti_xO_3$, and (3) a semi-conducting pyrochlore $Y_2Mo_2O_7$. Intriguingly, despite their distinct microscopic Hamiltonians, all of them exhibit unconventional glassy behaviors in the TRM measurements, that are weak and broad shoulder-like memory effects as in the prototype spin jam compounds SCGO/BCGO, starkly contrasting the strong and dip-like memory effects observed in the canonical spin glass such as *Cu*Mn2%. Interestingly, all the data can be well reproduced by a modified stretched exponential function of $\left\{1 - \exp\left(-\left(\frac{t_w}{\tau}\right)^{1-n}\right)\right\}$. More importantly, all the densely populated magnets except $Y_2Mo_2O_7$ yield an exponent of $1 - n \approx 0.6(1)$. This value is different from $1 - n \approx 1/3$ of dilute magnetic alloys[28-30] that was ascribed to their hierarchical and fractal energy landscape[31-34], and is also different from $1 - n = 1$ of the conventional Debye relaxation expected for a crystal. Based on these results, we argue that the glass magnets can be categorized into two distinct classes with different relaxation behaviors characterized by the exponent: $1 - n \approx 1/3$ for glassy magnets with hierarchical energy landscape and $1 - n \approx 0.6(1)$ for the ones with non-hierarchical energy landscape.

The TRM measurement is the most effective way to probe the memory effects in detail as explained in Supplementary Information and as shown most recently in the comparative study[35] of SCGO/BCGO and the canonical spin glass *Cu*Mn2%. While a dip-like memory effect with clear rejuvenation was observed, as expected, in *Cu*Mn2%, a shoulder-like memory effect seen in SCGO/BCGO implies lack of rejuvenation. Figure 1 shows the TRM data obtained from five different compounds: (a) $Fe_{1.02}Se_{0.15}Te_{0.85}$, (b) $La_{1.96}Sr_{0.04}CuO_4$ (LSCO(x=0.04)), (c) $Li_2RhO_3$, (d) $Na_2Ir_{0.89}Ti_{0.11}O_3$, and (e) $Y_2Mo_2O_7$. These TRM data were taken after waiting at the waiting temperature $T_w \sim 0.7\, T_f$ for several different waiting times ranging from 1.5(5) min to maximally 100 hours. For all systems aging and memory effect appears, i.e., the magnetization decreases near $T_w$ when the measurements were performed after waiting. The memory effect gets enhanced as the waiting time, $t_w$, increases. Surprisingly, $Fe_{1.02}Se_{0.15}Te_{0.85}$ and LSCO(x=0.04) whose parent compounds, FeTe and $La_2CuO_4$, respectively, are long-range ordered state, i.e., spin solid, exhibit very weak memory effects. The memory effects in both systems are negligible for short waiting time $t_w \leq 6$ min. For $t_w \gtrsim 1$ hr, both systems show a very weak and broad shoulder appearing around $T_w$ (see Figs. 1a and 1b), regardless of how large $t_w$ is. For $Fe_{1.02}Se_{0.15}Te_{0.85}$, the memory effect even seems to saturate for $t_w \gtrsim 30$ hrs (Fig. 1a). Similar weak shoulder-like memory effects were also observed in the spin-orbit Mott insulators, $Li_2RhO_3$ and $Na_2Ir_{0.89}Ti_{0.11}O_3$ (see Fig. 1c and 1d, respectively). Note that, similarly to the two superconductivity-related systems, the two Kitaev-model-related systems also exhibit negligible memory effects for short waiting time $t_w \leq 6$ min.

The weak shoulder-like memory effects have been recently observed in frustrated magnets, SCGO and BCGO, that are in vicinity of spin liquid, and here we show that another frustrated magnet, $Y_2Mo_2O_7$, also exhibits similar features (Fig. 1e). These data clearly show that the weak shoulder-like memory effect is universal in these densely populated magnets, regardless of their magnetic interactions. It is in stark contrast to the canonical spin glass such as $Cu$Mn2%, where the memory effects in the magnetization curve were readily seen even for such short waiting times as $t_w = 1.5(5)$ min (see Fig. 3a, and Fig. S2 in the Supplementary Information), and the effects become sharp and strong, appearing as a large dip at $T_w$ for $t_w \geq 3$ hrs[35].

Figure 2a summarizes the $t_w$ dependence of the memory effect for the densely populated magnets along with the canonical spin glass $Cu$Mn2%. The relative change of the magnetization $\Delta M_{rel} = (M_{ref} - M)/M_{ref}$ induced by the aging, in which $M$ and $M_{ref}$ are the magnetizations with and without waiting, respectively, is plotted. Overall, it is clear that the memory effect is much weaker in densely populated magnets than in the canonical spin glass. Firstly, $\Delta M_{rel}$ of $t_w \geq 30$ hrs for all the densely populated magnets except $Y_2Mo_2O_7$ is smaller than $\Delta M_{rel}$ of $t_w = 6$ min for $Cu$Mn2%. Secondly, for $t_w \leq 6$ min, most of them show negligible memory effects as shown in Fig. 2a and Fig. 1. Thirdly, the memory effect of the spin jams except $Y_2Mo_2O_7$ seems to saturate for $t_w \geq 30$ hrs, while for $Cu$Mn2% it seems to keep increasing with increasing $t_w$ over the time period. It is interesting that the densely populated $Y_2Mo_2O_7$ exhibits both spin glass and spin jam behaviors. This is probably due to the fact that $Y_2Mo_2O_7$ is a semi-conductor evidenced by its resistivity of $\rho \sim 10^{-2}$ $\Omega \cdot$ cm at 300 K, and has an unquenched orbital degree of freedom[36,37]. As a result, $Y_2Mo_2O_7$ is not a typical frustrated antiferromagnet, which is manifested in the relatively small frustration index $f = \Theta_{CW}/T_f \cong 2.3$ that is two orders of magnitude smaller than that of SCGO.

In search of possible underlying scaling behavior, we have fitted the $t_w$ dependence of $\Delta M_{rel}$ to the following phenomenological function

$$\Delta M_{rel}(t_w) = A\left\{1 - \exp\left(-\left(\frac{t_w}{\tau}\right)^{1-n}\right)\right\}, \quad (1)$$

which is modified from the stretched exponential function that was proposed to describe relaxation phenomena in glassy systems[28-34]. The modification made here is to take into account the experimental observation that $\Delta M_{rel}$ seems to saturate for long waiting times. Here $A = \Delta M_{rel}(t_w \to \infty)$ is a measure of degree of aging, $\tau$ is a microscopic time scale for relaxation dynamics. A positive non-zero exponent $n$ would tell us how much the relaxation deviates from the conventional Debye behavior ($n = 0$). The exponent $1 - n$ can be related to critical exponents for the spin glass transition within the framework of a random cluster model[32,33]. For example, assuming that the growth of clusters involves no conserved mode, the droplet model predicts an exponent $1 - n = 1/2$[38]. The dashed lines in Figure 2a are the fits of the experimental data to Eq. (1) for all the materials. It is remarkable that the same phenomenological function, albeit with different parameters, reproduces all the data of both spin jams and spin glass over the wide range of the waiting time. This indicates that a universal scaling may be in play in the aging or relaxation phenomena of all glassy magnets, as shown in Fig. 2c.

The difference between the spin glass and spin jam is clearly manifested in different parameters in Fig. 2d. For spin glass $Cu$Mn2%, the exponent $1 - n \approx 1/3$ that deviates significantly from the conventional Debye behavior of $1 - n = 1$. This is consistent with the previous studies on several other dilute magnetic alloys such as $Cu$Mn1% and $Ag$Mn2.6%[28], NiMn23.5%[29], $Au_{90}Fe_{10}$[30]. This deviation observed in the spin glasses was ascribed to the underlying hierarchically constrained dynamics[31-34]. On the other hand, the densely populated glassy magnets, SCGO, $Fe_{1.02}Se_{0.15}Te_{0.85}$, LSCO, and the two spin-orbit Mott insulators, yield the exponent of $1 - n \approx 0.6(1)$, indicating a smaller deviation from the conventional Debye relaxation. This implies that their energy landscapes are not hierarchical as in the canonical spin glass. These are summarized in Fig. 2d in which the exponent $1-n$ is plotted as a function of the degree of aging, $A$. We note that there is a positive correlation between the deviation from the Debye limit and the degree of aging.

To further support the aforementioned scenario, we have performed the TRM measurements on $Cu_{1-x}Mn_x$ as a function of the Mn concentration, $x$. This series of compounds provides an excellent platform also to investigate how the spin glass is connected with the spin jam, and eventually magnetic ordered states. On one hand, $Cu_{1-x}Mn_x$ is a canonical spin glass for small $x$. On the other hand, pure Mn exhibits a long-range spin-density wave (SDW) order at low temperatures. The magnetic ground state of samples with large $x$ thus can be viewed as large domains of SDW order disrupted by non-magnetic Cu atoms, similar to that observed in the densely populated magnets such as $Fe_{1.02}Se_{0.15}Te_{0.85}$ and $La_{1.96}Sr_{0.04}CuO_4$.

As shown in Fig. 3a, 3b, 3c and 3d, for dilute alloys with small values of $x \lesssim 0.45$, the data exhibits prominent dip behaviors, i.e., the presence of rejuvenation. As $x$ increases further, the dip behavior is gradually replaced with the shoulder behaviors, i.e., lack of rejuvenation, similar to spin jam (see Fig. 3e and 3f). The crossover seems to occur at $x \sim 0.45$ that is close to the percolation threshold for a three-dimensional system[39]. Note the non-monotonic behavior of the degree of aging $A = \Delta M_{rel}(t_w \to \infty)$ that maximizes at $x \sim 0.15$. The initial growth of A for small $x$ is related to the increasing number of magnetic impurities, giving rise to a stronger magnetic signal. For very large $x$ where the system is in the spin jam regime, the degree of aging is expected to decrease as observed for $x = 0.75$ and 0.85 shown in Fig. 3e and 3f respectively. Thus, even though the exact value of $x$ for the maximal A is determined by the balancing between the exact nature of the magnetic interactions and the magnetic concentration, the maximum of A should occur most likely somewhere close to the middle of $x = 0$ and the percolation threshold, which is qualitatively consistent with the observed value of $x \sim 0.15$.

Surprisingly, regardless of $x$, $\Delta M_{rel}$ of $Cu_{1-x}Mn_x$ follows the same stretched exponential relaxation function, as shown in Fig. 2b, but with varying values of the exponent, $1-n$, from 0.34(1) for $x$=0.02 to 0.66(9) for $x$=0.85 (see Fig. 2d). And thus, all their $\Delta M_{rel}$ can be collapsed into a same function, once the waiting time is properly scaled, and it is so even with those of the densely populated glassy systems, as shown in Fig. 2c. The change in the exponent, $1-n$, as a function of $x$ clearly shows that the glassy state of the dilute magnetic alloy (for small $x$) is replaced by a glassy state for large $x$ similar to the one observed in the densely populated magnets (see Fig. 2d) Interesting, the crossover occurs as the magnetic concentration go beyond the percolation threshold[39]. This clear crossover phenomenon strongly indicates that there are two distinct glassy states: spin glass and spin jam.

Why do the densely populated systems exhibit the large exponent $1-n \approx 0.6(1)$ similar to the quantum-fluctuation-induced spin jam SCGO, compared to the canonical spin glass state of dilute magnetic alloys? A clue comes from neutron scattering studies; the magnetic structure factor, $I(Q)$, of all the densely populated magnets studied here exhibit prominent peaks that are centered at a non-zero momentum ($Q$) corresponding to short-range spin correlations, as those of the frustrated magnets SCGO[17,40] and BCGO[18]. This indicates that those systems have dominant antiferromagnetic interactions between localized spins and short-range spin correlations. For example, the cuprate[41,42] and iron chalcogenide[43,44] exhibit strong incommensurate peaks near the antiferromagnetic ordering wave vector of their parent compounds. As shown in Fig. 4b, the spin-orbit Mott insulator $Na_2Ir_{0.89}Ti_{0.11}O_3$ exhibit a prominent peak centered at $Q = 0.87(2)$ Å$^{-1}$. The common characteristics of the antiferromagnetic and short-range magnetic structure factor starkly contrast with the nearly featureless magnetic structure factor of the spin glass *Cu*Mn2%, as shown in Fig. 4d. In the dilute magnetic alloys such as *Cu*Mn2%, magnetic impurities interact among themselves through the Ruderman-Kittel-Kasuya-Yosida (RKKY) interactions that are mediated by the itinerant electrons. The RKKY interactions are long-ranged, and oscillate from ferromagnetic to antiferromagnetic as a function of the distance. As a result, the random distances among the magnetic moments lead to their random interactions that even change the sign, resulting in the featureless magnetic structure factor.

The featureless $I(Q)$ of $Cu$Mn2% is consistent with the real-space droplet model for spin glass[38,45] in which low-energy excitations are dominated by connected spin clusters of arbitrary length scales. The real-space clusters or droplets correspond to the meta-stable ground states or local minima in the energy landscape. Their arbitrary length scales and random RKKY interactions yield a multitude of energy scales, resulting in the complex hierarchical fractal energy landscape[35,46-48]. As a consequence, the spin glass exhibits the observed strong dip-like memory effect. In contrast to the droplet model for spin glass, the clusters in spin jams are more uniform in size, as evidenced by the prominent peak of $I(Q)$. This feature, combined with the short-range exchange spin Hamiltonian, leads to a narrowly distributed energy scale, and the weak memory effect as observed in our susceptibility measurements.

The distinct nature of the two magnetic glass phases, spin glass and spin jam, also manifests in their characteristically different low energy excitations. The thermodynamic behavior of canonical spin glass at low temperatures is dominated by thermally active clusters or droplets, particularly those with a free energy less than or of the same order of $k_B T$ where $k_B$ is the Boltzmann constant. The fact that there is a finite density of clusters with limiting zero free energy naturally leads to the linear-$T$ specific heat[38,49], which is a signature of canonical spin glass. On the other hand, the low-energy excitations in spin jam are the Halperin-Saslow (HS) spin waves with finite spin stiffness over large length scales (often larger than the typical cluster sizes)[50-52]. These gapless HS modes exhibit a linear dispersion relation and are the source of a $T^2$ dependence of the specific heat for a two-dimensional system. Indeed, such $T^2$ behavior has been observed in the glass phase of SCGO[14], $Li_2RhO_3$[8,9], and doped $Na_2IrO_3$[10].

The memory effect measurements provide crucial information about the nature of relaxation dynamics in different magnetic states, which allows us to classify the semi-classical magnetic glassy materials as shown in Fig. 5. At the lower left corner of the triangle lies the spin solid that is realized in densely populated semi-classical magnetic materials with small disorder and weak frustration that order long-range at low temperatures with Debye relaxation. The typical energy landscape associated with spin solid is a smooth vase with a global minimum. At the lower right corner of the triangle lies the spin glass that is realized in dilute magnetic alloys with random magnetic interactions. Its typical energy landscape is dominated by hierarchical meta-stable states that correspond to spin clusters of arbitrary length scales in real space, exhibiting hierarchical rugged funnels and fractal geometry, and the observed strong deviation from the conventional Debye relaxation. Finally, at the top corner is the new magnetic state dubbed spin jam that encompasses many densely populated compounds with short-range exchange magnetic interactions, disorder and frustration. Disorder can be either extrinsic as in LSCO, FeTeSe and $Na_2Ir_{1-x}Ti_xO_3$, or intrinsic due to quantum fluctuations as in SCGO and BCGO [15,16]. One salient feature of the spin jam, represented by a nonhierarchical energy landscape with a wide and nearly flat but rough bottom, is the lack of widely distributed energy and time scales. This in turn leads to a significantly weaker memory effect and the relaxation exponent that is closer to the Debye exponent than that of the spin glass, as observed in our experiments. Remarkably, the canonical spin glass $Cu_{1-x}Mn_x$ with small $x$ crosses over to the spin jam state when the magnetic concentration x increases beyond the percolation threshold.

Our classification of a wide range of semi-classical glassy magnets based on nonequilibrium relaxation dynamics to two distinct states has implication to other non-magnetic structural glasses. Indeed, recent studies have found two distinct low frequency modes in structural glass: one related to a hierarchical energy landscape and the other related to jamming[53-55]. The rather distinct aging and memory behaviors observed in the spin glass and jam might also shed light on the relationship between nonequilibrium dynamics and connectivity among elementary interacting agents in networks and socio-economic systems[56].

## Acknowledgments

Work at University of Virginia by S.H.L. and A.S. was supported by US National Science Foundation (NSF) Grant DMR-1404994 and Oak Ridge National Laboratory, respectively. The work at Tohoku University was partly supported by Grants-in-Aid for Scientific Research (24224009, 23244068, and 15H05883) from MEXT of Japan. The work at University of Tennessee was supported by the US National Science Foundation (NSF) grant DMR-1350002.

## Author contributions

S-H. L. designed the research. J.Y., N. K., R. S., H. D. Z., S. W. and K. Y. synthesized the samples. A. S., T. J. S. and M. T. performed the memory effect measurements. A. S., D. Z., G. E., S. O. D. and D. A. T. performed the neutron scattering experiments. G-W. C. and S-H. L. wrote the paper.

## Additional Information

**Competing financial interests:** The authors have no competing interests as defined by Nature Publishing Group, or other interests that might be perceived to influence the results and/or discussion reported in this paper.

## References

1. Mydosh, J.A. *Spin Glasses: An Experimental Introduction* (Taylor & Francis, 1995).
2. Cannella, V. & Mydosh, J.A. Magnetic Ordering in Gold-Iron Alloys, *Phys Rev B* **6**:4220 (1972).

3. Anderson, P. W. Spin Glass, *Phys Today* **41**:9 (1988).

4. Dupuis, V., Bert, F., Bouchaud, J.-P., Hammann, J., Ladieu, F., Parker, D. & Vincent, E. Aging, rejuvenation, memory phenomena in spin glasses, *Pramana*, **64**(6), 1109-1119 (2005).

5. Wakimoto, S., Birgeneau, R. J., Kastner, M. A., Lee, Y. S., Erwin, R., Gehring, P. M., Lee, S. H., Fujita, M., Yamada, K., Endoh, Y., Hirota, K., & Shirane, G. (2000) Direct observation of a one-dimensional static spin modulation in insulating $La_{1.95}Sr_{0.05}CuO_4$, *Phys Rev B* **61**(5):3699-3706.

6. Matsuda, M., Fujita, M., Yamada, K., Birgeneau, R. J., Kastner, M. A., Hiraka, H., Endoh, Y., Wakimoto, S., & Shirane, G. Static and dynamic spin correlations in the spin-glass phase of slightly doped $La_{2-x}Sr_xCuO_4$, *Phys Rev B* **62**(13):9148-9154 (2000).

7. Katayama, N., Ji, S., Louca, D., Lee, S. H., Fujita, M., Sato, T. J., Wen, J., Xu, Z., Gu, G., Xu, G., Lin, Z., Enoki, M., Chang, S., Yamada, K., & Tranquada, J. M. Investigation of the Spin-Glass Regime between the Antiferromagnetic and Superconducting Phases in $Fe_{1+y}Se_xTe_{1-x}$, *J Phys Soc Jpn* **79**:113702 (2010).

8. Luo, Y., Cao, C., Si, B., Li, Y., Bao, J., Guo, H., Yang, X., Shen, C., Feng, C., Dai, J., Cao, G., & Xu, Z. $Li_2RhO_3$: A spin-glassy relativistic Mott insulator, *Phys Rev B* **87**(16):161121(R) (2013).

9. Khuntia, P., Manni, S., Foronda, F. R., Lancaster, T., Blundell, S. J., Gegenwart, P., & Baenitz, M. Local Magnetism and Spin Dynamics of the Frustrated Honeycomb Rhodate $Li_2RhO_3$, *arXiv*:**1512**.04904 (2015).

10. Manni, S., Tokiwa, S. Y., & Gegenwart, P. Effect of nonmagnetic dilution in the honeycomb-lattice iridates $Na_2IrO_3$ and $Li_2IrO_3$, *Phys Rev B* **89**: 241102 (2014).

11. Gingras, M. J. P., Stager, C. V., Raju, N. P., Gaulin, B. D., & Greedan, J. E. Static Critical Behavior of the Spin-Freezing Transition in the Geometrically Frustrated Pyrochlore Antiferromagnet $Y_2Mo_2O_7$, *Phys Rev Lett* **78**(5): 947-950 (1997).

12. Mamiya, H. et al. Slow dynamics in the geometrically frustrated magnet $ZnFe_2O_4$: Universal features of aging phenomena in spin glasses, *Phys Rev B* **90**:014440 (2014).

13. Obradors, X., et al. Magnetic frustration and lattice dimensionality in $SrCr_8Ga_4O_{19}$, *Solid State Commun* **65**(3):189–192 (1988).

14. Ramirez, A. P., Espinosa, G. P. & Cooper, A. S. Strong frustration and dilution-enhanced order in a quasi-2D spin glass, *Phys Rev Lett* **64**(17): 2070-2073 (1990).

15. Iida, K., Lee, S. H. & Cheong, S. W. Coexisting Order and Disorder Hidden in a Quasi-Two-Dimensional Frustrated Magnet, *Phys Rev Lett* **108**:217207 (2012).

16. Klich, I., Lee, S. H. & Iida, K. Glassiness and exotic entropy scaling induced by quantum fluctuations in a disorder-free frustrated magnet, *Nat Comm* **5**: 3497 (2014).

17. Yang, J. et al. Spin jam induced by quantum fluctuations in a frustrated magnet, *Proc Nat Acad Sci USA* **112**:11519 (2015).

18. Yang, J., Samarakoon, A. M., Hong, K. W., Copley, J. R. D., Huang, Q., Tennant, A., Sato, T. J., & Lee, S. H. Glassy Behavior and Isolated Spin Dimers in a New Frustrated Magnet BaCr$_{9p}$Ga$_{12-9p}$O$_{19}$, *J Phys Soc Jpn* **85**: 094712 (2016).

19. Aeppli, G., Hayden, S. M., Mook, H. A., Fisk, Z., Cheong, S. W., Rytz, D., Remeika, J. P., Espinosa, G. P., & Cooper, A. S. Magnetic dynamics of La$_2$CuO$_4$ and La$_{2-x}$Ba$_x$CuO$_4$, *Phys Rev Lett* **62**:2052-2055 (1989).

20. Coldea, R., Hayden, S. M., Aeppli, G., Perring, T. G., Frost, C. D., Mason, T. E., Cheong, S. W., & Fisk, Z. Spin Waves and Electronic Interactions in La$_2$CuO$_4$, *Phys Rev Lett* **86**(23):5377-5380 (2001).

21. Bao, W., Qiu, Y., Huang, Q., Green, M. A., Zajdel, P., Fitzsimmons, M. R., Zhernenkov, M., Chang, S., Fang, M., Qian, B., Vehstedt, E. L., Yang, J., Pham, H. M., Spinu, L., & Mao, Z. Q. Tunable $(\delta\pi, \delta\pi)$-Type Antiferromagnetic order in α-Fe(Te,Se) Superconductors, *Phys Rev Lett* **102**:247001 (2009).

22. Dai, P., Hu, J., & Dagotto, E. Magnetism and its microscopic origin in iron-based high-temperature superconductors, *Nature Phys* **8**:709 (2012).

23. Katukuri, Y. M., Nishimoto, S., Rousochatzakis, I., Stoll, H., van den Brink, J., & Hozoi, L. Strong magnetic frustration and anti-site disorder causing spin-glass behavior in honeycomb Li$_2$RhO$_3$, *Scientific Reports* **5**:14718 (2015).

24. Choi, S. K., Coldea, R., Kolmogorov, A. N., Lancaster, T., Mazin, I. I., Blundell, S. J., Radaelli, P. G., Singh, Y., Gegenwart, P., Choi, K. R., Cheong, S. W., Baker, P. J., Stock, C., & Taylor, J. Spin Waves and Revised Crystal Structure of Honeycomb Iridate Na$_2$IrO$_3$, *Phys Rev Lett* **108**:127204 (2012).

25. Mamiya, H. & Nimori, S., Memory effects in Heisenberg spin glasses: Spontaneous restoration of the original spin configuration rather than preservation in a frozen state, *J Appl Phys* **111**: 07E147 (2012).

26. Ladieu, F., Bert, F., Dupuis, V., Vincent, E. & Hammann, J. The relative influences of disorder and of frustration on the glassy dynamics in magnetic systems, *J Phys: Condens Matter* **16**: S735-S741 (2004).

27. Dupuis, V. et al. Aging and memory properties of topologically frustrated magnets, *J Appl Phys* **91**: 8384 (2002).

28. Chamberlin, R. V., Mozurkewich, G., & Orbach, R. Time decay of the remanent magnetization in spin-glasses, *Phys Rev Lett* **52**, 867-870 (1984).

29. Roshko, R. M. & Ruan, M. Thermoremanent relaxation in a reentrant *Ni*Mn ferromagnet close to the tricritical point, *J Mag Mag Mat* **104** 1613-1614 (1992).

30. Mitchler, P., Roshko, R. M., & Ruan, W. Crossover from equilibrium to nonequilibrium dynamics in a reentrant *Au*Fe ferromagnet, *Journal de Physique I* **2**, 2299-2309 (1992).

31. Palmer, R. G., Stein, D. L., Abrahams, E. & Anderson, P. W. Models of hierarchically constrained dynamics for glassy relaxation, *Phys Rev Lett* **53**(10):958 (1984).

32. Continentino, M. A. & Malozemoff, A. P. Dynamic scaling and the field-dependent critical line in a fractal cluster model of spin glasses, *Phy Rev B* **33**(5), 3591 (1986).

33. Campbell, I. A. Critical exponents of spin-galss systems, *Phys Rev B* **37**(16), 9800 (1988).

34. Ogielski, A. T. Dynamics of three-dimensional Ising spin glasses in thermal equilibrium, *Phys Rev B* **32**(11), 7384 (1985).

35. Samarakoon, A. M., Sato, T. J., Chen, T., Chern, G. W., Yang, J., Klich, I., Sinclair, R., Zhou, H., & Lee, S. H. Aging, memory, and nonhierarchical energy landscape of spin jam, *Proc Nat Acad Sci USA* **113**:11806 (2016).

36. Silverstein, H. J., et al. Liquidlike correlations in single-crystalline $Y_2Mo_2O_7$: An unconventional spin glass, *Phys Rev B* **89**: 054433 (2014).

37. Shinaoka, H., Motome, Y., Miyake, T. & Ishibashi, S. Spin-orbital frustration in molybdenum pyrochlores $A_2Mo_2O_7$ (A= rare earth), *Phys Rev B* **88**(17):174422 (2013).

38. Fisher, D. & Huse, D. Equilibrium behavior of the spin-glass ordered phase, *Phys Rev B* **38**:386-411 (1988).

39. Henley, C. L. Effective Hamiltonians and dilution effects in Kagome and related anti-ferromagnets. Can J Phys 79(11–12):1307–1321 (2001).

40. Lee, S. H., Broholm, C., Aeppli, G., Perring, T. G., Hessen, B., & Taylor, A. Isolated Spin Pairs and Two-Dimensional Magnetism in $SrCr_{9p}Ga_{12-9p}O_{19}$, *Phys Rev Lett* **76**(23):4424-4427 (1996).

41. Cheong, S. W., Aeppli, G., Mason, T. E., Mook, H., Hayden, S. M., Canfield, P. C., Fisk, Z., Clausen, K. N., & Martinez, J. L. Incommensurate magnetic fluctuations in $La_{2-x}Sr_xCuO_4$, *Phys Rev Lett* **67**(13): 1791-1794 (1991).

42. Emery, V. J., Kivelson, S. A., & Tranquada, J. M. Stripe phases in high-temperature superconductors, *Proc Natl Acad Sci USA* **96**:8814-8817 (1999).

43. Lee, S. H., Xu, G., Ku, W., Wen, J. S., Lee, C. C., Katayama, N., Xu, Z. J., Ji, S., Lin, Z. W., Gu, G. D. & Yang, H. B. Coupling of spin and orbital excitations in the iron-based superconductor $FeSe_{0.5}Te_{0.5}$, *Phys Rev B* **81**(22) 220502(R) (2010).

44. Liu, T. J., Hu, J., Qian, B., Fobes, D., Mao, Z. Q., Bao, W., Reehuis, M., Kimber, S. A. J., Prokeš, K., Matas, S. & Argyriou, D. N. From $(\pi, 0)$ magnetic order to superconductivity with $(\pi, \pi)$ magnetic resonance in $Fe_{1.02}Te_{1-x}Se_x$, *Nat Mater* **9**:716-720 (2010).

45. Bouchaud, J. P., Dupuis, V., Hammann, J., & Vincent, E. Separation of time and length scales in spin glasses: Temperature as a microscope, *Phys Rev B* **65**: 024439 (2001).

46. Lederman, M., Orbach, R., Hammann, J. M., Ocio, M., & Vincent, E. Dynamics in spin glasses, *Phys Rev B* **44**(14):7403-7412 (1991).

47. Fontanari, J. F. & Stadler, P. F. Fractal geometry of spin-glass models, *J Phys A: Math Gen* **35**(7):1509-1516 (2002).

48. Charbonneau, P., Kurchan, J., Parisi, G., Urbani, P., & Zamponi, F. Fractal free energy landscapes in structural glasses, *Nat Comm* **5**:3725 (2014).

49. Anderson, P. W., Halperin, B. I., & Varma, C. M. Anomalous low-temperature thermal properties of glasses and spin glasses, *Philos Mag* **25**:1 (1972).

50. Halperin, B. I. & Saslow, W. M. Hydrodynamic theory of spin waves in spin glasses and other systems with noncollinear spin orientations, *Phys Rev B* **16**(5): 2154-2162 (1977).

51. Sachdev, S. Kagomé- and triangular-lattice Heisenberg antiferromagnets: Ordering from quantum fluctuations and quantum-disordered ground states with unconfined bosonic spinons, *Phys Rev B Condens Matter* **45**(21):12377–12396 (1992).

52. Podolsky, D. & Kim, Y. B. Halperin-Saslow modes as the origin of the low-temperature anomaly in $NiGa_2S_4$, *Phys Rev B* **79**(14):140402 (2009).

53. Franz, S., Parisi, G., Urbani, P. & Zamponi, F. Universal spectrum of normal modes in low-temperature glasses, *Proc Nat Acad Sci USA* **112**:14539 (2015).

54. Chen, K., Ellenbroek, W. G., Zhang, Z., Chen, D. T., Yunker, P. J., Henkes, S., Brito, C., Dauchot, O., Van Saarloos, W., Liu, A. J. & Yodh, A. G. Low-frequency vibrations of soft colloidal glasses, *Phys Rev Lett* **105**(2), 025501 (2010).

55. Liu, A. J. & Nagel, S. R. The jamming transition and the marginally jammed solid, *Annu Rev Condens Matter Phys* **1**(1) 347-369 (2010).

56. Albert, R. & Barabasi, A. -L. Statistical mechanics of complex networks, *Rev Mod Phys* **74**:47-97 (2002).

**Figure captions**

**Fig. 1**. **Memory Effect as a function of waiting time.** Bulk susceptibility, $\chi_{DC} = M/H$, where $M$ and $H$ are magnetization and applied magnetic field strength, respectively, obtained from (a) $Fe_{1.02}Se_{0.15}Te_{0.85}$ (b) $La_{1.96}Sr_{0.04}CuO_4$, (c) $Li_2RhO_3$ (d) $Na_2Ir_{0.89}Ti_{0.11}O_3$ and (e) $Y_2Mo_2O_7$, with H = 3 Oe. Symbols and lines with different colors indicate the data taken with different waiting times, $t_w$, ranging from zero to 100hrs, at $T_w/T_f \sim 0.7$ where $T_w$ and $T_f$ are the waiting and the freezing temperature, respectively. For Fe$_{1.02}$Se$_{0.15}$Te$_{0.85}$, the Curie-Weiss Temperature $\theta_{cw}$ was estimated by fitting its high-T susceptibility data as shown in Fig. S1A in the Supplementary Information. For La$_{1.96}$Sr$_{0.04}$CuO$_4$, the high-T susceptibility does not follow the simple Curie-Weiss law (see Fig. S1b in Supplementary Information). In order to show how strong the magnetic interactions are in LSCO, we quote the coupling constants of the parent compound La$_2$CuO$_4$ that were experimentally determined by inelastic neutron scattering (ref. 20); the antiferromagnetic nearest-neighbor $J \approx 104\ meV$ and the ferromagnetic next-nearest-neighbor $J' \approx -18\ meV$. $\theta_{cw}$ for Li$_2$RhO$_3$, Na$_2$Ir$_{0.89}$Ti$_{0.11}$O$_3$ and Y$_2$Mo$_2$O$_7$ were taken from ref. 9, 24 and 36, respectively.

**Fig. 2. Summarizing the memory effect.** From the data shown in (a) Fig. 1 and (b) Fig.3, the aging effect was quantified for the eleven systems by plotting the relative change of the magnetization $\Delta M_{rel} = (M_{ref} - M)/M_{ref}$ where $M_{ref}$ is the magnetization without waiting, and it was plotted as a function of $t_w$ in a log scale. The aging effects of a spin jam prototype, SrCr$_{9p}$Ga$_{12-9p}$O$_{19}$ (SCGO(p=0.97)), and a spin glass prototype $Cu$Mn2% were taken from Ref. 35, except the $t_w = 1.5(5)\ min$ data are new (see Fig. S2 in the Supplementary Information), and are also plotted here for comparison. Each set of $\Delta M_{rel}(t_w)$ for each sample shown in panels (a) and (b) was fitted to the modified stretched exponential function, Eq. (1). After the fitting, in (c) $-\log(1 - \Delta M_{rel}/A)$ was plotted as a function of $(t_w/\tau)^{1-n}$ in a log-log scale. (d) The degree of aging, $A$, and the inverse exponent, $1/(1-n)$, obtained for all the samples are plotted against each other.

**Fig. 3. Memory Effect of Cu -x at. % Mn samples as a function of waiting time.** Bulk susceptibility, $\chi_{DC} = M/H$, where $M$ and $H$ are magnetization and applied magnetic field strength, respectively, obtained from Cu$_{1-x}$Mn$_x$ with (a) $x$=0.02, (b) $x$=0.15, (c) $x$=0.30, (d) $x$=0.45, (e) $x$=0.75 and (f) $x$=0.85, with H = 3 Oe. Symbols and lines with different colors indicate the data taken with different waiting times, $t_w$, ranging from zero to 100hrs, at $T_w/T_f \sim 0.7$ where $T_w$ and $T_f$ are the waiting and the freezing temperature, respectively.

**Fig. 4. Neutron scattering measurements.** (a) T-dependence, $I_{elas}(T)$, and (b) Q-dependence, $I_{elas}(Q)$, of elastic magnetic neutron scattering intensity obtained from Na$_2$Ir$_{0.89}$Ti$_{0.11}$O$_3$. The measurements were done at the Cold Neutron Chopper Spectrometer (CNCS) at the Spallation Neutron Source (SNS). (c) T-dependence, $I_{elas}(T)$, and (d) Q-dependence, $I_{elas}(Q)$, of elastic magnetic neutron scattering intensity obtained from the magnetic alloy $Cu$Mn2%. The measurements were done at the Backscattering Spectrometer (BASIS) at SNS. For both $I_{elas}(Q)$ in (b) and (d), the non-magnetic background was determined from the data above the freezing temperature and subtracted from the base temperature data. The black solid line in (b) is the fit of the magnetic peak centered at $Q = 0.87\ Å^{-1}$ to a simple Gaussian, while the line in (d) is a guide to eyes. The red horizontal bar at the center of the peak in (b) represents the instrument Q-resolution, $dQ \approx 0.06$ Å$^{-1}$, that was determined by fitting a nearby Bragg peak centered at 1.2 Å$^{-1}$.

**Fig. 5. Schematic phase diagram.** Classification of semi-classical magnetic states into three distinct phases, spin solid, spin jam, and spin glass, was made based on the memory effect.

Fig. 1

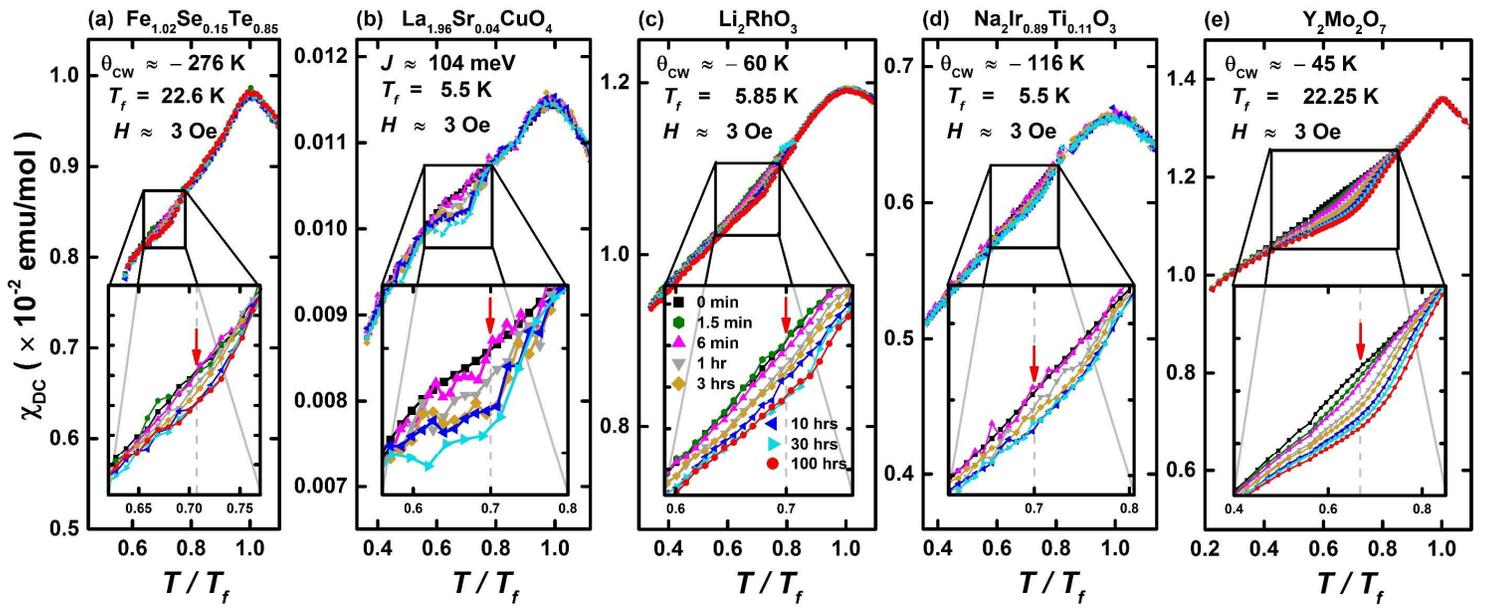

Fig. 2

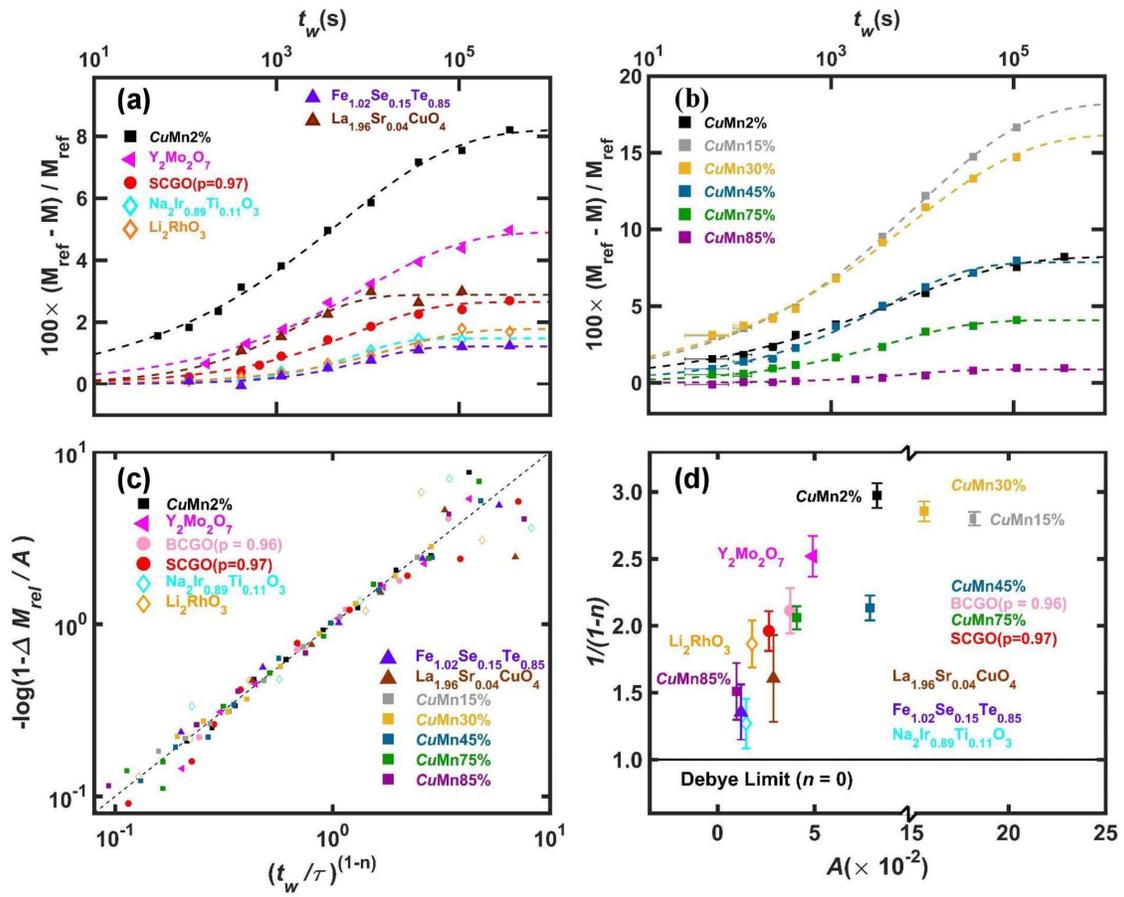

Fig. 3

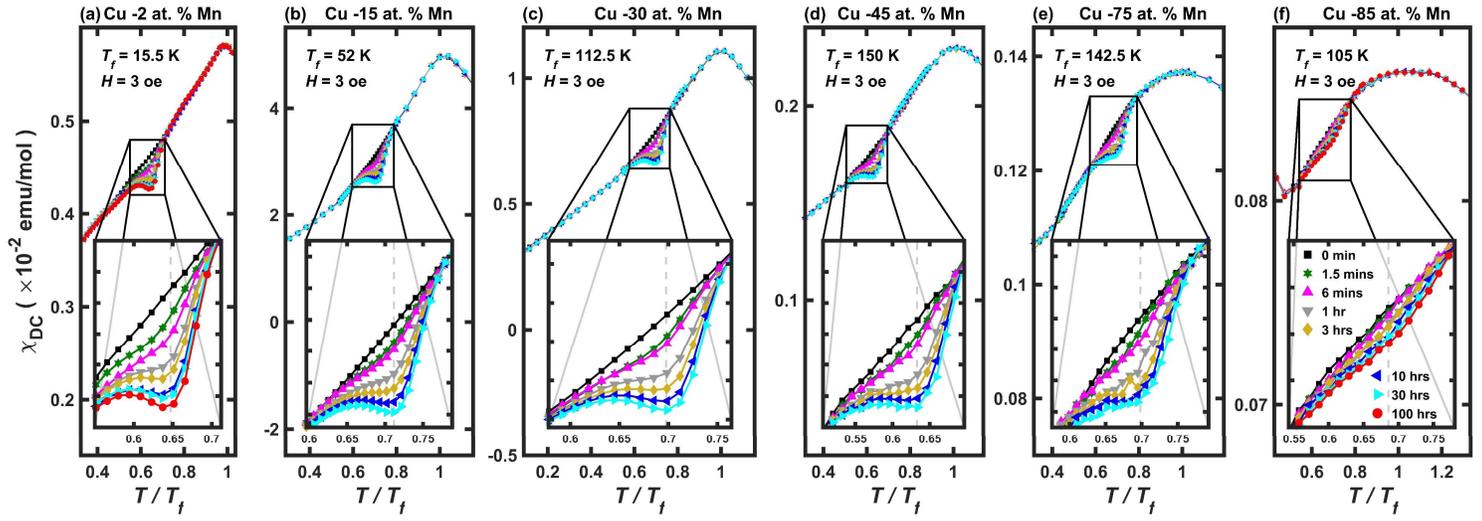

Fig. 4

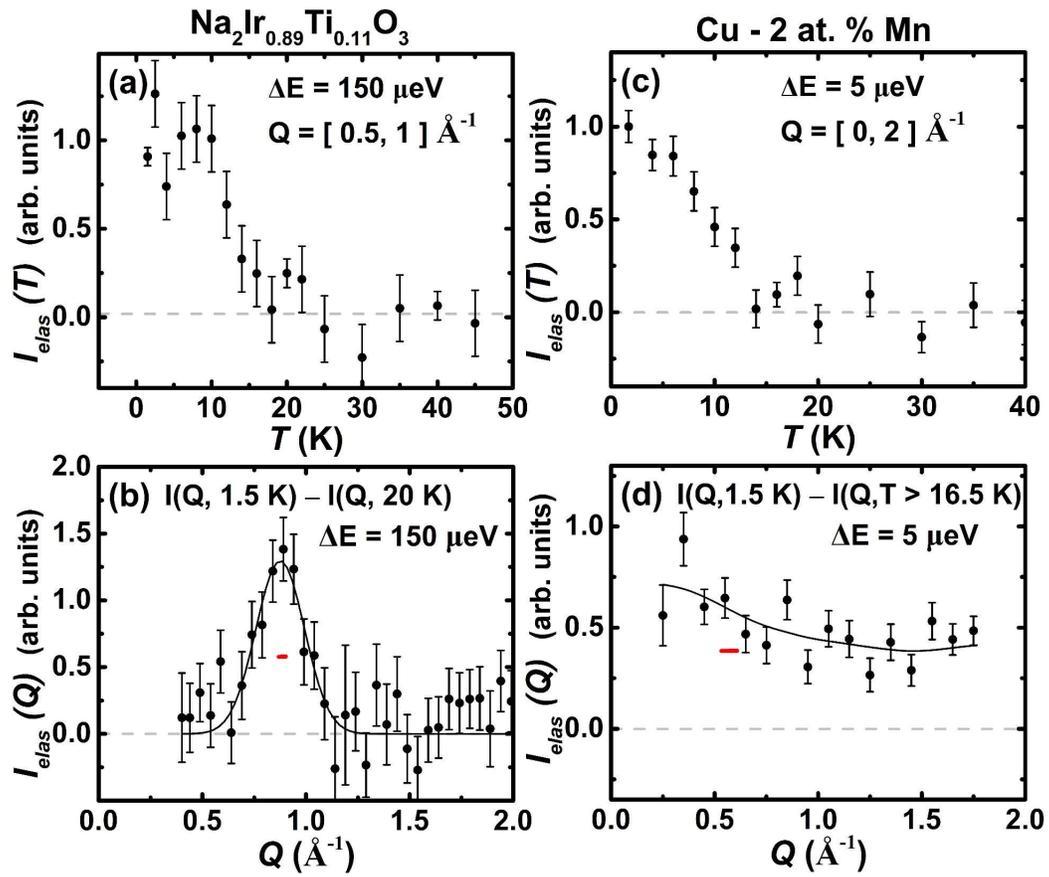

Fig. 5

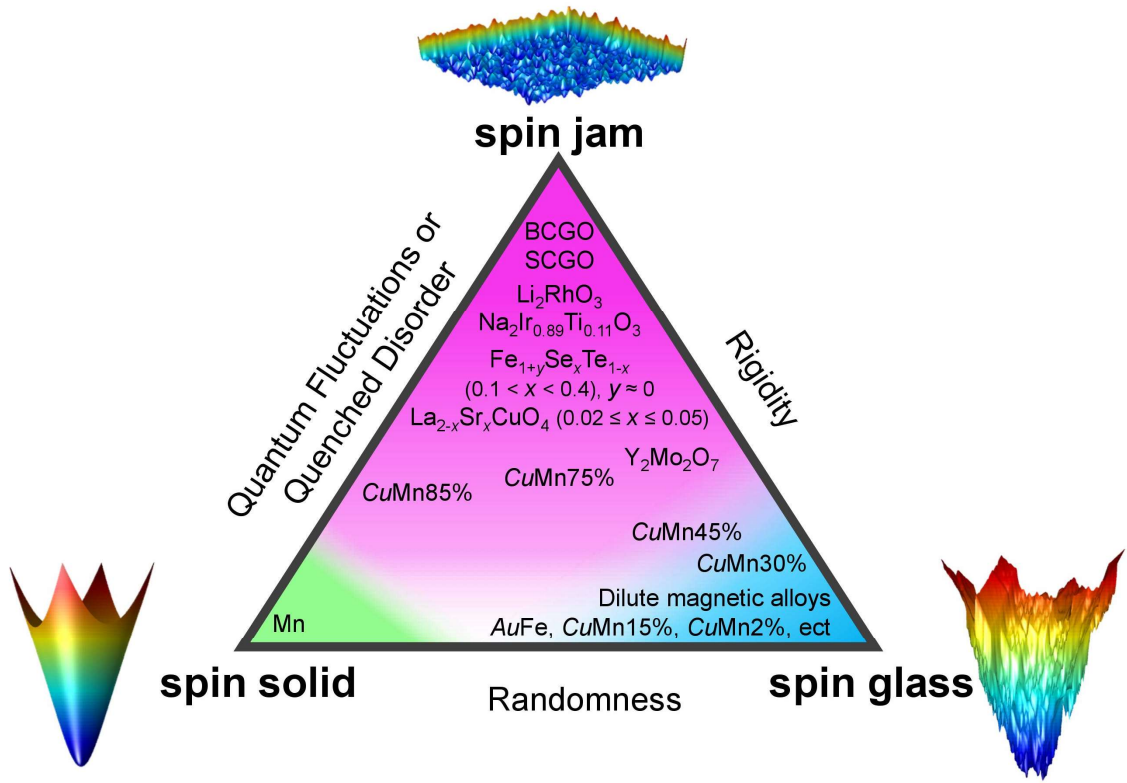

# Supplementary Information

# Scaling of Memories in Glassy Magnets

A. M. Samarakoon, M. Takahashi, D. Zhang, J. Yang, N. Katayama, R. Sinclair, H. D. Zhou, S. O. Diallo, G. Ehler, D. A. Tennant, S. Wakimoto, K. Yamada, G-W. Chern, T. J. Sato, S.-H. Lee

**This PDF file includes:**

I. Experimental procedure of the Thermo-Remanent Magnetization (TRM)
II. Neutron Scattering Methods

Figures S1-S3

## I. Experimental procedure of the Thermo-Remanent Magnetization (TRM)

The Figure S1 shows the DC susceptibility data obtained from $Fe_{1.02}Se_{0.15}Te_{0.85}$ and $La_{1.96}Sr_{0.04}CuO_4$ that exhibit their glassy transitions at low temperatures.

The Thermo-Remanent Magnetization (TRM) data, shown in Figure 1 and Figure 3 of the main text, and Figures S2 and S3 of the Supplementary Information, were collected using the following procedure. First, each sample was cooled down from well above the freezing temperature, $T_f$, to base temperature with a single stop for a period of time, $t_w$, at an intermediate temperature $T_w$ below $T_f$ under zero field. Once cooled down to base temperature, the thermo-remanent magnetization is measured by applying a small field of a few gauss upon heating at a constant rate. For all the measurements reported in this paper, we used a Superconducting Quantum Inference Device (SQUID) magnetometer, Quantum Design MPMS-XL5 equipped with the ultra-low-field option together with the environmental magnetic shield. Since it is necessary to have zero-field conditions at the sample during the cooling process including the waiting at an intermediate temperature, the remanent magnetic field at the sample position was measured by the instrument's fluxgate, and has been eliminated by introducing a compensating field using non-superconducting DC coil to get the remaining uncompensated magnetic field less than 0.005 G at the sample position. After that, a small DC magnetic field of 3 G was generated by the DC non-superconducting coil and applied to the sample during the TRM measurements.

Figure S3 shows that for the spin jam systems the memory effect with $t_w = 10$ hrs is maximal when the waiting temperature $T_w \sim 0.7\, T_f$ and it becomes weaker for other values of $T_w$ over a wide range of $T_w$.

## II. Neutron Scattering Methods

For the neutron scattering study of *Cu*Mn2%, the Backscattering Spectrometer (BASIS) at Spallation Neutron Source (SNS) was used. A 10 g polycrystalline sample of *Cu*Mn2% was sealed in a standard aluminum (Al) and was cooled in a standard liquid He-4 cryostat. During the measurements, the wavelength of scattered neutrons was fixed to be 6.2 Å by silicon analyzer crystals, yielding an elastic energy resolution of ~4 $\mu eV$. For the neutron scattering study of $Na_2Ir_{0.89}Ti_{0.11}O_3$, the Cold Neutron Chopper Spectrometer (CNCS) at SNS. A 2.3 g polycrystalline sample of $Na_2Ir_{0.89}Ti_{0.11}O_3$ was sealed in an Al annular can with thickness of 1 mm to reduce the neutron absorption by Ir, and was placed inside a standard liquid He-4 cryostat that can go down to 1.4 K. The wavelength of incident neutrons was fixed to be $\lambda = 5$ Å, yielding an elastic energy resolution of ~ 70 $\mu eV$. Elastic magnetic Neutron scattering intensity $I_{elas}(Q,T) = \int_{-\omega_0}^{\omega_0} I(\omega, Q, T) d\omega$, where $\omega_0$ is the instrument's elastic energy resolution has been determined by subtracting measurements done well above the freezing temperature $T_f$.

## Figure Captions

**Figure S1:** High-Temperature bulk susceptibility (black) and inverse susceptibility (red) respectively, obtained from (a) $Fe_{1.02}Se_{0.15}Te_{0.85}$ and (b) $La_{1.96}Sr_{0.04}CuO_4$. The data above 120 K of $Fe_{1.02}Se_{0.15}Te_{0.85}$ has been fitted to the Curie-Weiss law (red dash line) and the estimated Curie-Weiss temperature is -265.5(8) K. The measurements have done under magnetic fields of 0.01 T and 0.1 T respectively.

**Figure S2:** Bulk susceptibility, $\chi_{DC} = M/H$, where $M$ and $H$ are magnetization and applied magnetic field strength, obtained from (a) $Cu$Mn2% and (b) $SrCr_{9p}Ga_{12-9p}O_{19}$ (p=0.97) with $H = 3\ Oe$. The $t_w = 1.5(5)$ min data is new while all other data for $t_w \geq 6$ min are taken from Ref. 35.

**Figure S3.** Temperature Dependence of memory effect. $\chi_{DC}$ and $(M_{ref} - M)/M_{ref}$ measured for (a) $Fe_{1.02}Se_{0.15}Te_{0.85}$ (b) $La_{1.96}Sr_{0.04}CuO_4$, (c) $Li_2RhO_3$ (d) $Na_2Ir_{0.89}Ti_{0.11}O_3$ and (e) $Y_2Mo_2O_7$, with $t_w = 10\ hrs$, at various waiting temperatures.

Fig. S1

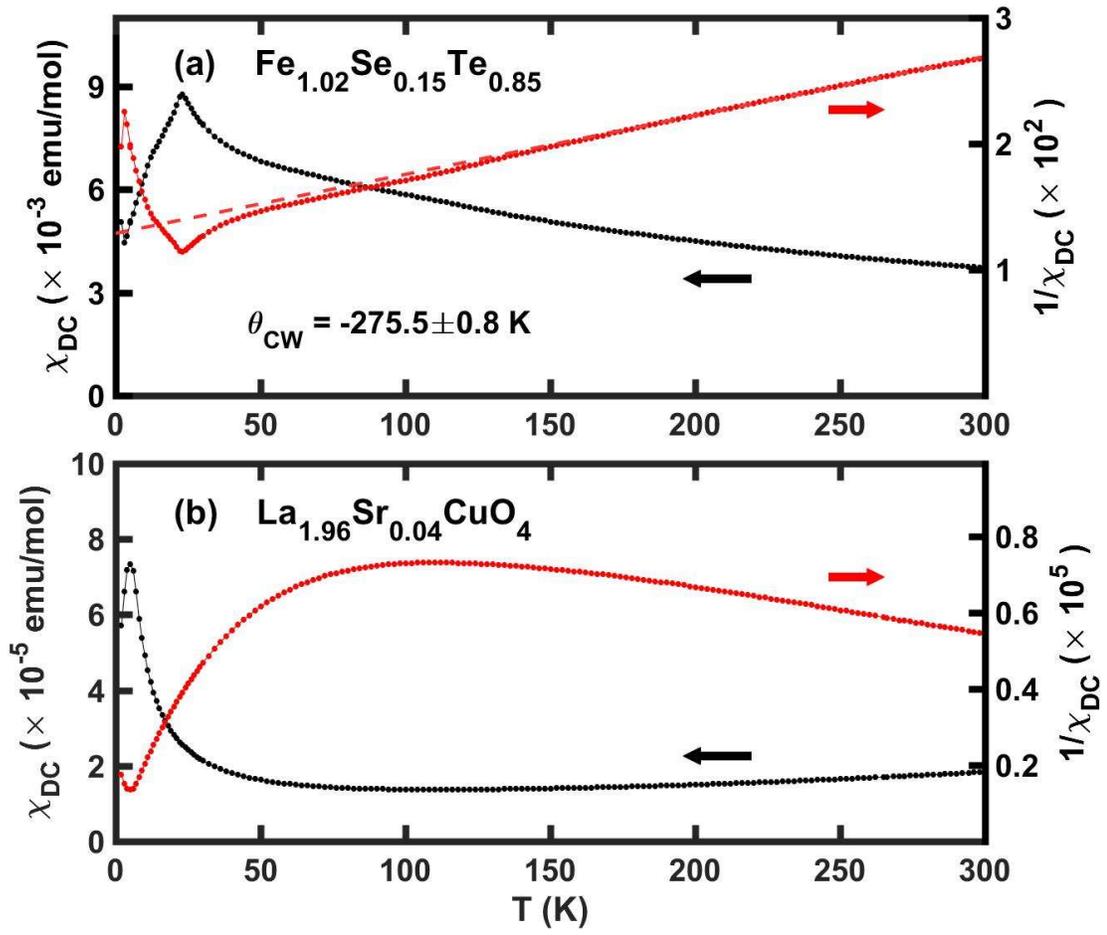

Fig. S2

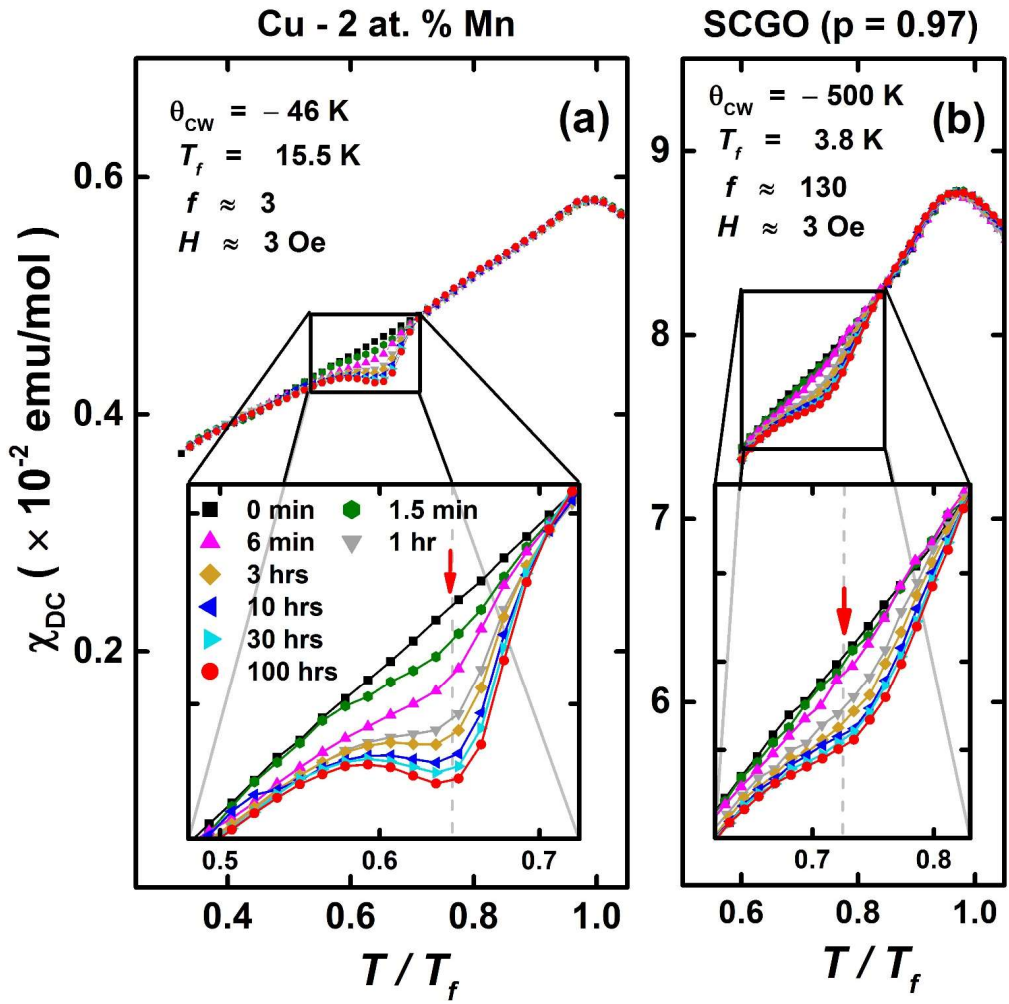

Fig. S3

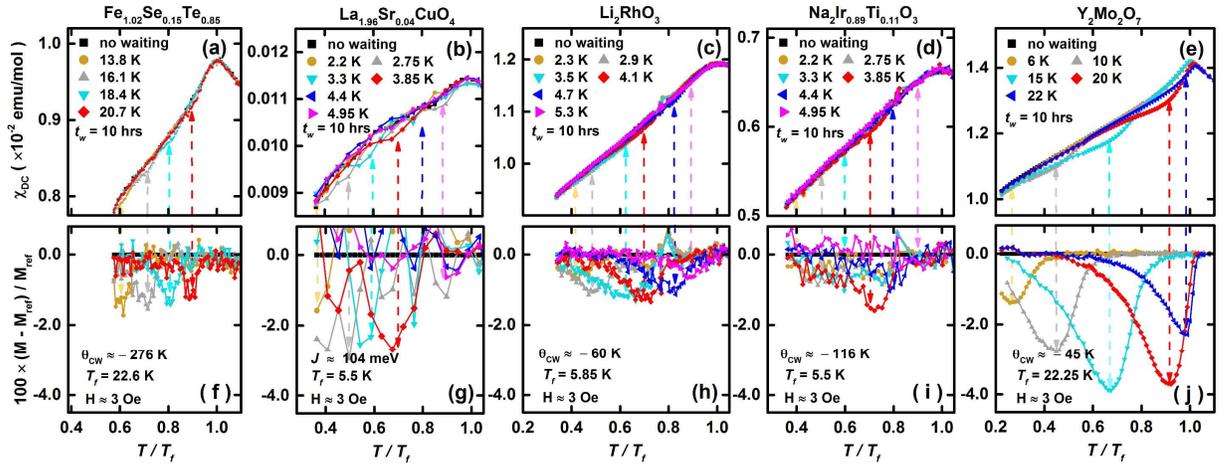